\documentclass[aip,jcp,reprint]{revtex4-1}

\usepackage{amsmath, amsfonts,braket,graphicx,color}

\global\long\def\p{\prime}

\global\long\def\tr{\mathrm{tr}}

\global\long\def\im{\imath}

\newcommand{\dg} {{\dagger}}
\newcommand{\pd} {{\phantom\dagger}}

\newcommand{\ci}[1] {{c_{#1}^{\pd}}}
\newcommand{\cid}[1] {c_{#1}^\dg}

\newcommand{\ql}{\mathcal{L}}
\newcommand{\qs}{\mathcal{S}}
\newcommand{\qr}{\mathcal{R}}
\newcommand{\qe}{\mathcal{E}}
\newcommand{\qi}{\mathcal{I}}
\newcommand{\qu}{\mathcal{U}}
\newcommand{\qc}{\mathcal{C}}

\newcommand{\rs}{\bar{\rho}}
\newcommand{\EO}{E}
\newcommand{\EC}{\mathcal{J}}

\newcommand\trick[1]{}

\begin{document}

\title{Master Equations for Electron Transport: The Limits of the Markovian Limit}

\author{Justin E. Elenewski}

\affiliation{Center for Nanoscale Science and Technology, National Institute of
Standards and Technology, Gaithersburg, MD 20899, USA}

\affiliation{Maryland Nanocenter, University of Maryland, College Park, MD 20742, USA}

\author{Daniel Gruss}

\affiliation{Center for Nanoscale Science and Technology, National Institute of
Standards and Technology, Gaithersburg, MD 20899, USA}

\affiliation{Maryland Nanocenter, University of Maryland, College Park, MD 20742, USA}

\author{Michael Zwolak}

\email{mpz@nist.gov}

\affiliation{Center for Nanoscale Science and Technology, National Institute of
Standards and Technology, Gaithersburg, MD 20899, USA}

\begin{abstract}
Master equations are increasingly popular for the simulation of time--dependent electronic transport in nanoscale devices. Several recent Markovian approaches use ``extended reservoirs'' -- explicit degrees of freedom associated with the electrodes -- distinguishing them from many previous classes of master equations. Starting from a Lindblad equation, we develop a common foundation for these approaches. Due to the incorporation of explicit electrode states, these methods do not require a large bias or even ``true Markovianity'' of the reservoirs. Nonetheless, their predictions are only physically relevant when the Markovian relaxation is weaker than the thermal broadening and when the extended reservoirs are ``sufficiently large,'' in a sense that we quantify. These considerations hold despite complete positivity and respect for Pauli exclusion at any relaxation strength.
\end{abstract}
\maketitle

Nanoscale electronics have made inroads into a diverse range of applications, from tunneling-based DNA sequencing \cite{ZwolakDiVentra2005, LagerqvistZwolakDiVentra2006, ZwolakDiVentra2008, BrantonDeamerMarzialiEtAl2008, ChangHuangHeEtAl2010, TsutsuiTaniguchiYokotaEtAl2010, HuangHeChangEtAl2010, OhshiroMatsubaraTsutsuiEtAl2012} to high--performance microelectronics \cite{buttiker2006mesoscopic, gabelli2006violation, Lansbergen2008, Morello2010, Tan2010, Pierre2010, Fuechsle2012, Shulaker2013, Perrin2015, trasobares201617}.  The theoretical description of these devices is complicated by strong environmental effects, which profoundly influence electronic transport and lead to behavior beyond the static Landauer formalism. While a formally exact solution for such time-dependent transport exists, it requires the use of computationally demanding two--time Green's functions \cite{Meir1992, Jauho1994}, which are impractical for many applications.  The description of sensing devices also necessitates an accounting of atomic fluctuations and unknown structural details, complicating their simulation.

One major goal is the development of a theoretical framework that can circumvent these limitations while remaining versatile enough to augment contemporary electronic structure methods.  Quantum master equations and density--matrix propagation afford such an approach \cite{Frensley1985, Frensley1990, Knezevic2013}, and encompass a diversity of well--established schemes that lie largely in the Markovian limit \cite{Fischetti1998, Fischetti1999, Mizuta1991, Knezevic2008, Schaller2008, Rosati2014, Rosati2017}, with notable non--Markovian extensions \cite{Schaller2009,Zedler2009,Vaz2010}.  While these methods have recently come to the forefront, their conceptual history dates to the early work of Kohn and Luttinger \cite{Kohn1957}, with subsequent developments that defined a device and its contacts as an open quantum system \cite{Frensley1985,Frensley1990}.

\begin{figure}
\includegraphics[width=\columnwidth]{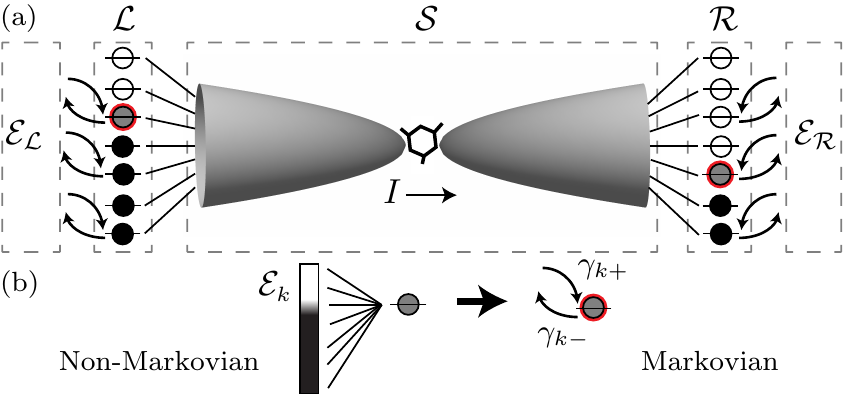}
\caption{Schematic of electronic transport. (a) ``Extended reservoir'' regions $\ql$ and $\qr$ drive a current $I$ through the system $\qs$. Since the extended reservoirs are finite, external environments $\qe_\ql$ and $\qe_\qr$ act to relax them back to equilibrium, maintaining a true steady state. (b) Each extended reservoir state $k$ is dressed by an infinite environment $\qe_k$ in equilibrium (yielding non-Markovian dynamics) or with properly balanced injection, $\gamma_{k+}$, and depletion, $\gamma_{k-}$, rates directing the state back to equilibrium (yielding Markovian dynamics).} \label{fig:schematic}
\end{figure}

We focus on a specific Markovian master equation,

\begin{equation} \label{eq:phenomEQ}
\overbrace{\dot{\rs} = -\frac{\im}{\hbar} [\bar{H}, \rs] - \gamma (\rs - \rs_0)}
         ^{\text{Single Particle Density Matrix}}
\Leftrightarrow
\overbrace{\dot{\qc} = -\frac{\im}{\hbar} [\bar{H}, \qc] - \gamma (\qc - \qc_0)}
         ^{\text{Single Particle Correlation Matrix}},
\end{equation}
in a particular context where both {\em explicit} and {\em implicit} extended reservoirs are present (introduced below) \cite{Gruss2016}. This master equation corresponds to a ``relaxation approximation'' where the system relaxes to $\rs_0$ ($\qc_0$)~\footnote{These are not always true density/correlation matrices.} at some rate $\gamma$.   We have written this expression in terms of both the single-particle density matrix, $\rs$, and the correlation matrix, $\qc_{kl} = \tr \, [ \cid{k} \ci{l} \, \rho ]$, since many recent works have focused on noninteracting systems. The single-particle Hamiltonian $\bar{H}$ is defined through $H=\sum_{k,l} \bar{H}_{lk} \cid{k} \ci{l}$, where $\ci{k}$ ($\cid{k}$) are the creation (annihilation) operators for the state $k$\footnote{It is convenient to have $\bar{H}_{lk}$ instead of $\bar{H}_{kl}$ in $H$ to give the commutator in Eq.~\eqref{eq:phenomEQ} rather than one with a transpose.}. We reserve $\rho$ and $H$ for the full, many-body density matrix and Hamiltonian \footnote{The quantities $\rs$ and $\qc$ give all the information required to reconstruct the full density matrix for noninteracting systems. If, though, one takes $\rs$ as trace one, then the total number of particles is also required to reconstruct $\rho$, but $\qc$ always contains the number of particles}. Strictly speaking, $\rho$, $\qc$, and $H$ are defined for an arbitrary number of particles, allowing for fluctuations during time evolution.  A current will flow when the reservoir component of $\rs_0$ is ``polarized'' by different chemical potentials.  This approach was applied to mean field electrons in Refs.~\onlinecite{Sanchez2006, Subotnik2009} by casting $\rs_0$ in a specific form, while other works employ an alternative $\rs_0$ that includes coherences between the device and reservoirs \cite{Chen2014, Zelovich2014, Zelovich2015, Zelovich2016, Hod2016, Morzan2017, Zelovich2017}.  

While related relaxation-type approximations have a lengthy history \cite{Kohn1957,Frensley1985, Frensley1990, Fischetti1998, Fischetti1999, Mizuta1991}, this specific dual--reservoir setup is new and foundational to a family of promising real--time simulation methods~\cite{Gruss2016, Sanchez2006, Subotnik2009, Chen2014, Zelovich2014, Zelovich2015, Dzhioev2011, Ajisaka2012, Ajisaka2013, Dast2014, Ajisaka2015, Mahajan2016, Zanoci2016, Hod2016, Zelovich2016, Zelovich2017, Morzan2017}. Here, we provide a rigorous justification to this setup, leading to both a well--defined domain of applicability and a connection between different variants of the formalism. More explicitly, this yields a mathematical rationale for their use in arbitrary systems  (e.g., in terms of reservoir sizes, many-body interactions, etc.) and identifies relevant physical limitations, laying the foundation for future applications and implementations.

One issue with Eq.~\eqref{eq:phenomEQ} is that -- while it is Markovian -- it is not in the standard Lindblad form \cite{Lindblad1976, Gorini1976}. As such, it is not obviously positive, and may yield both unphysical results and negative probabilities under certain conditions. For noninteracting electrons, the use of Eq.~\eqref{eq:phenomEQ} has been shown to be positive for asymptotically large reservoirs~\cite{Hod2016}. However, rather than start from Eq.~\eqref{eq:phenomEQ}, we would like an expression that is already in Lindblad form, which will allow us to guarantee complete positivity. 

We begin by examining the model depicted in Fig.~\ref{fig:schematic} and analyzed in Ref.~\onlinecite{Gruss2016}, where two electronic reservoirs, left ($\ql$) and right ($\qr$), drive current through a device $\qs$ that contains the system of interest (for instance, a nanoscale junction and its electronic leads). The reservoir regions are finite and explicitly part of the simulation. We term these ``extended reservoirs'' to distinguish from the typical assumption that they are infinite and implicit \cite{Meir1992, Jauho1994}. In order to have a true steady state, {\em implicit} environments $\qe_{\ql(\qr)}$ are introduced to relax $\ql$ ($\qr$) to their equilibrium distributions -- the notion of equilibrium is central to the use of these Master equations. The Hamiltonian for this setup is
\begin{equation} \label{eq:theHam}
H = H_\qs + H_\ql + H_\qr + H_\qi,
\end{equation}
where $H_\qs$ is the Hamiltonian for $\qs$, potentially including many-body interactions, $H_{\ql(\qr)} = \sum_{k\in \ql(R)} \hbar \omega_k \cid{k} \ci{k}$ are the ``extended reservoir'' Hamiltonians, and $H_\qi = \sum_{k\in \ql\qr} \sum_{i\in \qs} (\hbar v_{ki} \cid{k} \ci{i} + \text{h.c.})$ is the interaction that couples them. The index $k$ includes all labels (electronic state, spin, reservoir), while $\omega_k$ and $v_{ki}$ denote the level and hopping frequencies.

The $\ql \qs \qr$ system is open. Under the influence of $\qe_{\ql(\qr)}$, its dynamics is given by the Markovian master equation
\begin{align} \label{eq:fullMaster}
\dot{\rho} = - \frac{\im}{\hbar} [H, \rho]
    &+ \sum_k \gamma_{k+} \left( \cid{k} \rho \ci{k}
        - \frac{1}{2} \left \{ \ci{k} \cid{k}, \rho\right \}\right) \notag \\
    &+ \sum_k \gamma_{k-} \left( \ci{k} \rho \cid{k}
        - \frac{1}{2} \left \{ \cid{k} \ci{k}, \rho \right \} \right)
\end{align}
for the {\em full, many-body density matrix} $\rho$. The first term on the right is the Hamiltonian evolution of $\rho$ under $H$ and the second (third) term reflects particle injection (depletion) into the state $k$ at a rate $\gamma_{k+}$ ($\gamma_{k-}$). To ensure that the reservoirs relax to equilibrium -- a pseudo-equilibrium, as we will see -- in the absence of $\qs$, $\gamma_{k+} \equiv \gamma f^\alpha (\omega_k)$ and $\gamma_{k-} \equiv \gamma [1 - f^\alpha (\omega_k)]$, where $f^\alpha (\omega_k)$ is the Fermi-Dirac distribution in the $\alpha \in \{\ql,\qr\}$ reservoir and  with $\gamma$ nonzero only for reservoir states.  We assume a general case where each reservoir may be at a different chemical potential or temperature. This specific master equation has appeared in previous efforts~\cite{Gruss2016, Dzhioev2011, Ajisaka2012, Ajisaka2013, Dast2014, Ajisaka2015,Mahajan2016, Zanoci2016}. In particular, Ref.~\onlinecite{Gruss2016} derives the closed form solution for both the interacting and noninteracting cases, as well as those for the related non-Markovian problem.

To connect Eq.~\eqref{eq:fullMaster} to noninteracting approaches \cite{Sanchez2006, Subotnik2009,Chen2014, Zelovich2014, Zelovich2015, Zelovich2016, Hod2016, Morzan2017, Zelovich2017}, we first differentiate the single--particle correlation matrix $\qc$, employ Eq.~\eqref{eq:fullMaster}, and use that $\tr \left( \cid{k} \ci{l} [H,\rho] \right) =  [\bar{H},\qc]_{kl}$, yielding
\begin{equation} \label{eq:C}
\dot{\qc} = -\im [\bar{H},\qc]/\hbar + R[\qc].
\end{equation}
The quantity $R[\qc]$ is the relaxation,
\begin{align}
(R[\qc])_{kl}
    &= \gamma_{k+} \delta_{kl} -
        \frac{\qc_{kl}}{2}\left(\gamma_{k+} + \gamma_{k-}
                                + \gamma_{l_+} +\gamma_{l-}\right) \notag \\
     &= \gamma \left[f_k^\alpha \delta_{kl} \delta_{k \in \ql\qr}
         - \frac{ \qc_{kl}}{2} (\delta_{k \in \ql\qr}
                                + \delta_{l \in \ql\qr}) \right] , \label{eq:Rkl}
\end{align}
where $\gamma_{i\pm}=0$ when $i \in \qs$, $\delta_{k \in \alpha}$ is 1 when $k\in\alpha$ (and zero otherwise), and $\delta_{kl}$ is the typical Kronecker delta.

Taking the block form
\begin{equation}
\qc = \left( \begin{array}{ccc}
    \qc_{\mathcal{L,L}} & \qc_{\mathcal{L,S}} & \qc_{\mathcal{L,R}} \\
    \qc_{\mathcal{S,L}} & \qc_{\mathcal{S,S}} & \qc_{\mathcal{S,R}} \\
    \qc_{\mathcal{R,L}} & \qc_{\mathcal{R,S}} & \qc_{\mathcal{R,R}}
\end{array} \right) ,
\end{equation}
where $\qc_{\alpha , \alpha^\p}$ are for a subset of states, i.e., in the regions $\alpha,\alpha^\p \in \{\ql, \qs, \qr\}$, the relaxation component becomes
\begin{equation} \label{eq:R}
R [\qc] = -\gamma \left(
\begin{array}{ccc}
    (\qc_{\mathcal{L,L}} - \qc_{0}^{\ql}) & \frac{1}{2} \qc_{\mathcal{L,S}}
        & \qc_\mathcal{{L,R}} \\
    \frac{1}{2} \qc_{\mathcal{S,L}} & 0 & \frac{1}{2} \qc_{\mathcal{S,R}} \\
    \qc_{\mathcal{R,L}} & \frac{1}{2} \qc_{\mathcal{R,S}}
        & (\qc_{\mathcal{R,R}} - \qc^\qr_{0})
\end{array}
\right) .
\end{equation}
The ``relaxed'' distributions are $\qc^{\alpha}_0 = \text{diag}[f^\alpha (\omega_k)]$. For simplicity, Eqs.~\eqref{eq:Rkl} and~\eqref{eq:R} are written in the single-particle eigenbasis of the decoupled $\ql$, $\qs$, and $\qr$ regions.

The equation of motion defined by Eq.~\eqref{eq:C} and Eq.~\eqref{eq:R}, is exactly that of Refs.~\onlinecite{Chen2014, Hod2016, Zelovich2014, Zelovich2015, Zelovich2016, Zelovich2017,Morzan2017}. Since the starting expression is in Lindblad form, this demonstrates that these prior approaches use a completely positive, trace--preserving master equation for the single-particle matrices. Our derivation shows that {\em these properties always hold}, including for finite reservoirs, as well as those that are asymptotically large~\footnote{The latter was shown in Ref.~\onlinecite{Hod2016}, where they call ``leads'' what we call ``extended reservoirs.'' \protect\trick.}. Moreover, by virtue of the use of creation/annihilation operators in Eq.~\eqref{eq:fullMaster}, Pauli exclusion is obeyed even though particle number is not conserved. Furthermore, if we make an approximation where the off--diagonal coherences are negligible, the phenomenological expression in Eq.~\eqref{eq:phenomEQ} using the $\qc_0$ from Refs.~\onlinecite{Sanchez2006, Subotnik2009} is recovered. This is not, however, guaranteed to preserve positivity.

{\em Too $V$, or not too $V$?} In the preceding discussion, we adopted a Lindblad equation from the outset. To take a more foundational perspective, we can use the Born--Markov approach \cite{Breuer2002} to derive this equation. In doing so, we see that there must be two {\em implicit} reservoirs with a high voltage between them.  While this might appear to nullify the use of Eq.~\eqref{eq:fullMaster}, we demonstrate that our approach is physically applicable. The following derivation is presented for a single extended reservoir state, which is sufficient for completeness as these states are separately relaxed.

Each extended reservoir state $k$ connects to an {\em implicit} reservoir, the environment $\qe_k$ [the $\qe_{\ql(\qr)}$ from Fig.~\ref{fig:schematic}(a) are composed of all $\qe_k$ for $k \in \ql (\qr)$]. The Hamiltonian for this dressed, extended reservoir state [Fig.~\ref{fig:schematic}(b)] is
\begin{align} \label{eq:Hamk}
H_k &= \hbar \omega_{k} \cid{k} \ci{k}
        + \sum_{l \in \qe_k} \hbar \omega_{l} \cid{l} \ci{l}
        + \sum_{l \in \qe_k} \hbar \nu_{l}
            \left( \cid{l} \ci{k} + \cid{k} \ci{l} \right) \notag \\
    &\equiv \hbar \omega_{k} \cid{k} \ci{k}
        + \sum_{l \in \qe_k} \hbar \omega_{l} \cid{l} \ci{l}
        +  H^\p ,
\end{align}
with $H^\p = \cid{k} \EO + \EO^\dg \ci{k}$ and $\EO= \sum_{l \in \qe_k} \hbar \nu_{l} \ci{l}$. In the following, we work in the interaction picture, $\EO(t) = \qu_{\qe_k}^\dg \EO \qu_{\qe_k}$ with $\qu_{\qe_k} = \exp(-\im H_{\qe_k} t / \hbar)$, so that $\EO(t) = \sum_l \hbar \nu_l \ci{l} \exp (-\im \omega_l t)$ [similarly, $\ci{k}(t) = \ci{k} \exp (-\im \omega_k t)$].

\begin{figure*}[t]
\includegraphics[width=\columnwidth]{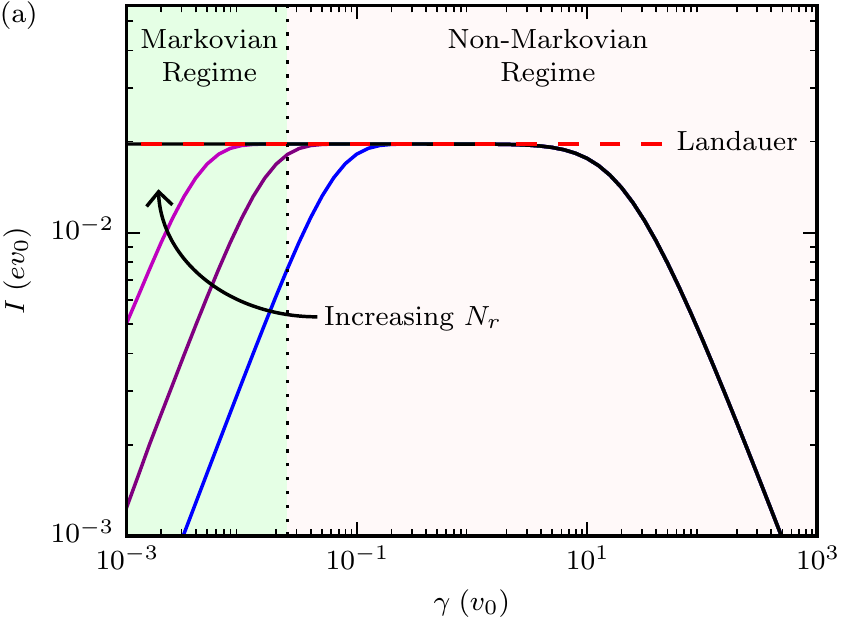}
\includegraphics[width=\columnwidth]{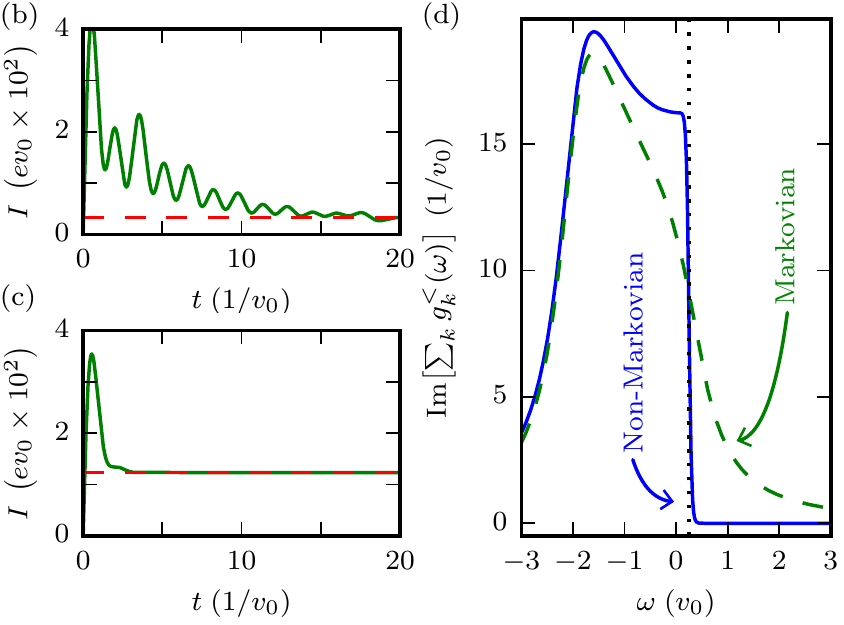}
\caption{(a) Steady-state current, $I$, as a function of the reservoir relaxation parameter, $\gamma$, Eq.~\eqref{eq:nonMarkSol}, with an applied bias of $V_{\ql (\qr)} = \pm v_0 \hbar / 4$ and an increasingly large number of explicit reservoir sites $N_r \in \{ 32, 128, 512 \}$. Here, we use a 1D model with an asymmetric shift: $w_k^{\ql (\qr)} = -2 v_0 \cos \left[ k \pi / (N_r + 1) \right] \pm v_0 / 10$ and $v_{ki} = v_0 \sqrt{2/(N_r + 1)} \sin \left[k \pi / (N_r + 1) \right]$ (with $k \in \{ 1, \dots, N_r \}$ for both $\ql$ and $\qr$), $k_B T = v_0 \hbar / 40$, and a single site at zero energy in $\qs$. In the limit of $N_r \rightarrow \infty$ (black line), the plateau extends to $\gamma=0$, yielding the Landauer/Mier-Wingreen current exactly (red dashed line). The dotted line demarcates where the Markovian master equation is valid, which can be reached numerically by increasing the number of reservoir sites. (b, c) Anomalous zero-bias ($V_{\ql (\qr)} = 0$) currents versus time $t$ for (b) a weak relaxation ($\gamma=0.1 v_0$) and (c) a strong relaxation ($\gamma=v_0$). The red dashed line indicates their limit as $t \to \infty$. In the non-Markovian case, however, there is zero steady state current. (d) Anomalous broadening of $g^<_k$ in $\ql$ for strong relaxation ($\gamma = v_0$) as shown by Eq.~\eqref{eq:nMtoM}. The solid blue line is the correct ``broadening before occupying'', while the dashed green line is the Markovian counterpart, which extends the occupation well beyond the Fermi level (black dotted line).  } \label{fig:curfig}
\end{figure*}

We begin with the Born--Markov master equation \footnote{In most scenarios, including ours, the first--order term vanishes, $\tr_{\qe_k}[H'(t), \rho(0)] = 0$},
\begin{equation} \label{eq:bmMaster}
\dot{\rho}_k (t) = -\frac{1}{\hbar^2} \int_0^\infty dt^\p \,
    \tr_{\qe_k}  \left[H'(t), \left[H'(t - t^\p),
        \rho_k(t) \otimes \rho_{\qe_k} \right]\right],
\end{equation}
where $\rho_{\qe_k}$ denotes the initial state of $\qe_k$. This approximation requires a weak coupling between $k$ and $\qe_k$ (essentially, second order perturbation theory) and the assumption of an uncorrelated, time-local composite state of the system and environment. The latter needs justification, which Eq.~\eqref{eq:Jp} will provide. Expanding the commutators and taking the trace in Eq.~\eqref{eq:bmMaster} gives 
\begin{align} \label{eq:rhok}
\dot{\rho}_k(t) =
    & \, (\gamma_{k-}/2) \left[\ci{k}  \rho_{k} \cid{k}
           - \cid{k} \ci{k} \rho_{k} \right]\mathclose{(t)}  
        + \text{h.c.} \notag \\
    &\, + (\gamma_{k+}/2) \left[\cid{k} \rho_{k} \ci{k} 
              - \ci{k}  \cid{k}  \rho_{k}  \right]\mathclose{(t)} 
       + \text{h.c.},
\end{align}
where the Hamiltonian component of the evolution will be recovered after returning from the interaction picture. The notation $[ \cdot ](t)$ indicates that all operators within the brackets are in the interaction picture. The relaxation $\gamma_{k\pm} = (2/\hbar^2) \int_0^\infty dt^\p \,  \EC_{\pm}(t^\p)  \exp (\mp\im \omega_k t^\p)$ is given in terms of the correlation functions $\EC_+(t^\p) = \tr_{\qe_k} \left[\EO^\dg (t^\p) \EO  \rho_{\qe_k} \right]$ and $\EC_-(t^\p)  =  \tr_{\qe_k} \left[\EO(t^\p) \EO^\dg \rho_{\qe_k} \right]$. An ideal Markovian environment exhibits only time-local correlations,
\begin{equation} \label{eq:Jp}
\EC_+(t^\p) = \int_{-\infty}^\infty d\omega \, J(\omega) f(\omega)
                e^{\im \omega t^\p}
            \equiv \delta(t^\p)\hbar^2\gamma_{k+} ,
\end{equation}
where a similar expression holds for $\EC_-(t^\p)$ [but with $f \to (1-f)$ and a change of sign in the exponent]. To obtain a $\delta$--function requires that the product $J(\omega) f(\omega)$ is constant, which can only be physically satisfied if the spectral function is flat, $J(\omega) \equiv \hbar^2 \gamma_{k+}/2 \pi$, and $f(\omega) \equiv 1$ for all $\omega$~\footnote{If the integral over $\omega$ is taken first, we take that $\int_{0}^\infty dt^\p \, \exp[\im \omega_k t^\p] \delta(t^\p) = 1/2$ (i.e., $\delta(t^\p)$ is treated as an even function).  If the integral over $t^\p$ is performed first, we use $\int_0^\infty dt^\p \, \exp[\im(\omega_k - \omega)t^\p] = \pi \delta(\omega_k - \omega) + \im \mathcal{P} \int_{-\infty}^\infty d\omega (\omega_k - \omega)^{-1}$, for which the principal value vanishes under Markovian conditions.}. In other words, the implicit reservoir is completely full. This is evident from the definition of a Markovian reservoir -- an environment that couples to the system equally at all frequency scales. The presence of the Fermi level breaks this symmetry. Thus, this level must lie at $\pm \infty$, as adopted in other efforts \cite{Zedler2009}.

Considering $\EC_-(t^\p)$, we find that $J(\omega) \equiv \hbar^2 \gamma_{k-}/2 \pi$ and $1-f(\omega) \equiv 1$. This implies that two distinct sets of states are required  to obtain Eq.~\eqref{eq:fullMaster}: In one set, the states are completely empty, acting only to deplete particles from $k$ [the first line of Eq.~\eqref{eq:rhok}]. In the other set, the states are completely full and thus they only inject particles into  $k$ [the second line of Eq.~\eqref{eq:rhok}]. References~\onlinecite{gurvitz1996microscopic, Gurvitz1998, sprekeler2004coulomb, Harbola2006} address this process when the implicit reservoirs connect directly to what we would call $\qs$, concluding that the equation of motion corresponds to a high bias $V$. In our approach, however, the implicit reservoirs are not connected directly to $\qs$, but rather indirectly through an intermediary -- the extended reservoirs.  The bias is thus wrapped into to the simulation and we do not require a high value of $V$. We will see this more explicitly below where we show that, when quantifying the errors of the Markovian Eq.~\eqref{eq:fullMaster} for steady states, no where does the bias show up, but rather only the temperature and extended reservoir size. The requirement for Markovianity may also be relaxed, since explicit reservoir states retain a memory up to time $1/\gamma$ of the dynamics. 

{\em True Limitations.} Equation~\eqref{eq:fullMaster} is completely positive, trace preserving, respects Pauli exclusion, and does not require a high $V$ for its use.  Nonetheless, these properties are not sufficient to ensure physically meaningful behavior.  To quantify this statement, we make use of the exact, closed form solution of Eq.~\eqref{eq:fullMaster} and its non--Markovian counterpart, both given in Ref.~\onlinecite{Gruss2016}. The latter also uses the model Hamiltonian from Eq.~\eqref{eq:Hamk}, however it does not require two distinct full and empty components of $\qe_k$ (nor a weak coupling between $k$ and $\qe_k$, a flat spectral function, or the wide--band limit). Rather, $\qe_k$ need only be infinite and in an equilibrium state described by the Fermi-Dirac distribution $f^{\ql (\qr)}$.

For this non-Markovian case, the current is
\begin{align} \label{eq:nonMarkSol}
I = -\frac{e}{2\pi} \int_{-\infty}^{\infty} d\omega \,
    &\left[ f^\ql(\omega) - f^\qr (\omega) \right] \\
    &\times \text{Tr} [\mathbf{\Gamma}^\ql(\omega) \mathbf{G}^r(\omega)
                       \mathbf{\Gamma}^{\qr}(\omega) \mathbf{G}^a (\omega)] ,
    \notag
\end{align}
where $\mathbf{G}^{r(a)} (\omega)$ are the -- potentially many-body -- retarded (advanced) Green's functions for $\qs$ [see Ref.~\onlinecite{Gruss2016} for the closed-form solution to the Markovian case, Eq.~\eqref{eq:fullMaster}]. The spectral densities of the couplings between $\qs$ and $\ql(\qr)$ are $\Gamma_{ij}^{ \ql(\qr)} (\omega) = \im \sum_{l \in \ql(\qr)} v_{j l} v_{l i} [g_{l}^r (\omega) - g_{l}^a (\omega)]$, defined in terms of the ``unperturbed'' -- but dressed -- extended reservoir state Green's functions, $g_k^{r(a)} (\omega) = [\omega - \omega_k \pm \im\gamma/2]^{-1}$. One may also obtain the lesser Green's function, $g_k^< (\omega) = -f^{\ql(\qr)}(\omega) [g_k^r(\omega) - g_k^a(\omega)]$.  For simplicity, we only address the wide--band limit (the more general case is in Ref.~\onlinecite{Gruss2016}). Notable in these expressions is the term $\gamma$, which accounts for relaxation and is key to subsequent discussion.  These results diverge from the  Meir--Wingreen formula \cite{Meir1992, Jauho1994} in the use of extended reservoirs as a finite-sized intermediary with relaxation.

While relaxation processes occur in real materials, these are not necessarily of the form in Eq.~\eqref{eq:Hamk}. Moreover, by taking Markovian relaxation as a further approximation, we cannot conclude that Eq.~\eqref{eq:fullMaster} will be physical for a given $\gamma$.  We thus interpret the relaxation as a control parameter, diligently chosen to obtain meaningful results from Eq.~\eqref{eq:fullMaster}. Numerical simulations [Fig.~\ref{fig:curfig}(a)] illustrate how the current in the full non-Markovian model behaves versus $\gamma$. There are three distinct regimes: A regime linear in $\gamma$, a plateau regime, and a $1/\gamma$ regime. In the intermediate plateau regime, the ``intrinsic conductance'' of the setup determines the current (for non--interacting systems, this would be the Landauer current). We note that the physics of the turnover versus $\gamma$ is analogous to Kramers' turnover for reaction rates in solution~\cite{Kramers1940}, and holds for thermal~\cite{VelizhaninSahuChienEtAl2015,chien2017,chien2017topological} as well as electronic transport~\cite{Gruss2016,Gruss2017}.

When simulating Markovian dynamics using Eq.~\eqref{eq:fullMaster}, instead of the non-Markovian problem, the three regimes are still present but a large $\gamma$ can result in non-zero currents at {\em zero bias} [Fig.~\ref{fig:curfig}(b,c)]. The origin of these currents is due to improper occupation of the extended reservoir levels.  Calculating the real-time correlation functions from  Eq.~\eqref{eq:fullMaster}, we see that the advanced and retarded Green's functions have a functional form that is identical to the non--Markovian case~\cite{Gruss2016}. In fact, the difference between the Markovian and non--Markovian limits is encapsulated by the replacement
\begin{equation} \label{eq:nMtoM}
g_k^< (\omega) =
    \overbrace{\frac{\im \gamma f^{\ql(\qr)} (\omega)}
                    {(\omega - \omega_k)^2 + \gamma^2 / 4}}
             ^{\text{Non-Markovian}}
\Rightarrow
    \overbrace{\frac{\im \gamma f^{\ql(\qr)} (\omega_k)}
                    {(\omega - \omega_k)^2 + \gamma^2 / 4}}
             ^{\text{Markovian}}.
\end{equation}
In the Markovian case, the extended reservoir state is being relaxed to the occupation $f^{\ql(\qr)} (\omega_k)$ and then broadened by $\gamma$. For the non--Markovian case, the $\gamma$--dressed state has the proper occupation $f^{\ql(\qr)} (\omega)$.  In other words, the non-Markovian case dresses then occupies and the Markovian case occupies then dresses. Figure~\ref{fig:curfig}(d) demonstrates that the Markovian limit gives an additional occupancy above the Fermi level, leading to zero--bias currents under certain conditions. Thus, the Markovian master equation, Eq.~\eqref{eq:fullMaster} or Eq.~\eqref{eq:C}, can yield unphysical behavior despite the fact that it is always completely positive and obeys Pauli exclusion. Another way to state the origin of this behavior is that the Markovian master equation relaxes the extended reservoir into a pseudo-equilibrium -- an equilibrium defined in terms of isolated extended reservoir states rather than those in the presence of the environment that provides the relaxation.

We can define precisely when the replacement in Eq.~\eqref{eq:nMtoM} yields a reasonable approximation, thus providing a satisfying quantification of Eq.~\eqref{eq:fullMaster}'s validity: So long as the $\gamma$-induced broadening is less than the thermal broadening $\gamma \ll k_B T/\hbar$, with temperature $T$ and $k_B$ the Boltzmann constant (or, in terms of timescales, $\gamma^{-1} \gg 25$~fs at room temperature), the Markovian limit accurately gives the steady state solutions~\cite{Gruss2016}. This is independent of {\em any} details of $\qs$ -- it may be interacting, non--interacting, have electron-phonon coupling, etc. The validity hinges on the replacement made in Eq.~\eqref{eq:nMtoM}, which, in turn, relies only on the fact that the reservoir states are non--interacting. This is generally a good approximation. While we do not qualify it here, the dynamics of interest will be correctly captured by Eq.~\eqref{eq:fullMaster} as long as they are faster than the relaxation, as the latter only cuts off behavior after a time $\gamma^{-1}$. These limits must be carefully enforced in order to ensure physically meaningful dynamics and steady state currents.

The considerations above lead to a natural estimate for the required number of extended reservoir states, $N_r$. So long as the extended reservoirs are sizable, one can take $\gamma \ll k_B T /\hbar$ and be within the plateau regime. A simple estimate is given by the turnover point between the linear and plateau regions. This generically occurs when $\gamma \approx W/N_r$, where $W$ is the bandwidth of the reservoirs, i.e., when $\gamma$ is on the order of the mode spacing in the extended reservoirs~\cite{Gruss2016,Gruss2017} (of course, inhomogeneity in the mode spacing can change this~\cite{Gruss2017}). Such behavior was recognized in early works that employed $\gamma > W/N_r$~\cite{Sanchez2006, Subotnik2009, Dzhioev2011}.

Putting these conditions together gives $N_r \approx \hbar W / k_B T$ (hence, $N_r$ should be very large at low temperature). A less stringent condition on $\gamma$ would only require the current is on the plateau, which often extends to relatively large values of $\gamma$. It is a mistake, however, to  conclude that an arbitrary, large value of $\gamma$ will be acceptable, even if this condition holds. {\em A sufficiently large $\gamma$ will improperly occupy high energy states, which -- when asymmetric reservoirs are present -- will give rise to unphysical zero-bias currents}~\footnote{When the reservoirs are symmetric, the forward and backward anomalous zero-bias currents cancel each other. However, this will still give anomalous correlations.}. Even though $\gamma$ plays the role of relaxation in the extended reservoirs, a small (or, in a sense, ``intermediary'') value is still necessary. 

We see that Markovian master equations can be a powerful tool for the simulation of electronic transport. Furthermore, the various approaches employed in the literature can be unified as equivalent expressions of Eq.~\eqref{eq:fullMaster} or some approximations thereof. These {\em extended reservoir}-based Markovian master equations do not require a large bias or even Markovianity. The true limit of the Markovian limit, Eq.~\eqref{eq:fullMaster}, is the requirement that $\gamma \ll k_B T /\hbar$ with an $N_r$ that is large enough to accommodate this slow relaxation and still yield the intrinsic conductance.

J. E. E. and D. G. acknowledge support under the Cooperative Research Agreement between the University of Maryland and the National Institute for Standards and Technology Center for Nanoscale Science and Technology, Award 70NANB14H209, through the University of Maryland.

\end{document}